\documentclass{PoS}

\usepackage{graphicx}
\graphicspath{{figures/}}

\title{The QCD vacuum wave functional and confinement in Coulomb gauge%
\addtocounter{footnote}{1}\thanks{This research was supported in part by the U.S.\ DOE under Grant No.\ DE-FG03-92ER40711 (J.G.), and by the Slovak Grant Agency for Science, Project VEGA No.\ 2/0070/09, by ERDF OP R\&D, Project CE QUTE ITMS~26240120009, and via CE SAS QUTE (\v{S}.O.)}}

\ShortTitle{The QCD vacuum wave functional and confinement in Coulomb gauge}

\author{Jeff Greensite\\
Physics and Astronomy Dept., San Francisco State University, San Francisco, CA 94132, USA\\
E-mail: \email{jgreensite@gmail.com}}

\author{\addtocounter{footnote}{-2}
				\speaker{\v{S}tefan Olejn{\'\i}k}\\
        Institute of Physics, Slovak Academy of Sciences, SK--845 11 Bratislava, Slovakia\\
        E-mail: \email{stefan.olejnik@savba.sk}}

\newcommand\FP{Faddeev--Popov\ }
\newcommand\YM{Yang--Mills\ }
\newcommand\tabentry[1]{\multicolumn{1}{c}{#1}}

\abstract{We report results on the Coulomb-gauge ghost propagator and the color-Coulomb potential computed in two lattice gauge-field ensembles: (1) configurations derived from our recently proposed \YM vacuum wave functional in $2+1$ dimensions, and (2) lattices generated by Monte Carlo simulations of the three-dimensional Euclidean SU(2) lattice gauge theory with the Wilson action. We observe remarkable agreement between the ghost propagators in both ensembles, but some differences in the potentials. Those originate from rare configurations with very small values of the lowest eigenvalue of the Coulomb-gauge \FP operator. If the same cuts on such exceptional configurations are applied in both ensembles, then the color-Coulomb potentials are also in reasonably good agreement.}

\FullConference{The XXVIII International Symposium on Lattice Field Theory\\
		 June 14--19, 2010\\
		 Villasimius, Sardinia, Italy}

\begin{document}

\section{Proposal for an approximate vacuum wave functional}
	Confinement is supposed to be encoded in properties of the vacuum of quantized non-abelian gauge theories. In the hamiltonian formulation in $D=d+1$ dimensions and temporal gauge, the vacuum wave functional satisfies the Schr\"odinger equation:
\begin{equation}
\displaystyle\int d^d x\left\lbrace-\frac{1}{2}
\frac{\delta^2}{\delta A_k^a(x)^2}+\frac{1}{4}F_{ij}^a(x)^2\right\rbrace\Psi_0[A]=E_0\Psi_0[A]
\end{equation}
together with the Gauss-law constraint:
\begin{equation}
\left(\delta^{ac}\partial_k+
g\epsilon^{abc}A_k^b\right)
\frac{\delta}{\delta A_k^c}\Psi_0[A]=0.
\end{equation}
At large distance scales one expects that the wave functional assumes the effective form:
\begin{equation}
\Psi_0^{\mathrm{eff}}[A]\approx\exp\left[-\mu\int d^dx\; F^a_{ij}(x)F^a_{ij}(x)\right].
\end{equation}
It has a property of \emph{dimensional reduction} \cite{Greensite:1979yn,Olesen:1981zp,Ambjorn:1984mb}: The computation of a spacelike loop in \linebreak $d+1$ dimensions reduces to the calculation of a Wilson loop in \YM theory in $d$ Euclidean dimensions. However, the true vacuum wave functional cannot be just like this -- it implies not only Wilson's area law, but also \textit{e.g.\/} exact Casimir scaling of higher-representation string tensions, which is not observed at asymptotic distances.

	Recently, we have proposed a simple approximation to the vacuum wave functional of the $(2+1)$-dimensional SU(2) \YM theory \cite{Greensite:2007ij}\footnote{A similar form, but without the -- crucial in our opinion -- subtraction of $\lambda_0$, was proposed by Samuel \cite{Samuel:1996bt}.}:
\begin{equation}\label{eq:GO-wf}
\Psi_0[A]{=}
{\cal{N}}\exp\left[-\frac{1}{2}\displaystyle\int d^2x\;d^2y\; 
B^a(x){\displaystyle
\left(\frac{1}{\sqrt{-{\cal D}^2-
\lambda_0+m^2}}\right)_{xy}^{ab}}B^b(y)\right],
\end{equation}
where $B^a(x)=F_{12}^a(x)$ denotes the color magnetic field strength, ${\cal D}_k[A]$ is the covariant derivative in the adjoint representation, ${\cal D}^2={\cal D}_k\cdot{\cal D}_k$ the covariant laplacian in the adjoint representation, $\lambda_0$ is the lowest eigenvalue of $(-{\cal D}^2)$, and $m$ is a constant (mass) parameter proportional to $g^2\sim 1/\beta$. The expression (\ref{eq:GO-wf}) is written in the continuum notation, but assumed to be properly defined on a lattice, where we choose
\begin{eqnarray}
\left({-{\cal D}^2}\right)^{ab}_{xy}&=&\displaystyle\sum_{k=1}^2 \left[2\delta^{ab}\delta_{xy}-
{\cal U}^{ab}_k(x)\delta_{y,x+\hat{k}}-{\cal U}^{\dagger ba}_k(x-\hat{k})\delta_{y,x-\hat{k}}\right],\\
{\cal U}^{ab}_k(x)&=&\frac{1}{2}\mbox{Tr}\left[\sigma^a U_k(x) \sigma^b U^\dagger_k(x)\right],
\end{eqnarray}
and $U_\mu(x)$ denotes the link matrix in the fundamental representation. 
 
\section{Arguments in favor of the proposed vacuum wave functional}

	In the original paper \cite{Greensite:2007ij}, we have supplied a series of arguments supporting the proposed form of the \YM vacuum wave functional:
\begin{enumerate}
\item
In the free-field limit ($g\to0$), $\Psi_0[A]$ becomes the well-known vacuum wave functional of electrodynamics.
\item
The proposed form is a good approximation to the true vacuum also for strong fields constant in space and varying only in time.
\item
If we divide the magnetic field strength $B(x)$ into ``fast'' and ``slow'' components, the part of the vacuum wave functional that depends on $B_\mathrm{slow}$ takes on the dimensional-reduction form. The fundamental string tension, at a given $\beta$, is then easily computed as $\sigma_\mathrm{F} = 3m/4\beta$.
\item
If one takes the mass $m$ in the wave functional as a free variational parameter and computes (approximately) the expectation value of the \YM hamiltonian, one finds that a non-zero (finite) value of $m$ is energetically preferred.
\end{enumerate}

\section{Lattice evidence}

	The above (analytic) hints are encouraging, but in no way sufficient to convince anybody that the simple vacuum wave functional, Eq.\ (\ref{eq:GO-wf}), is close to that of the true \YM ground state. To assess how good or bad the approximate state is, some numerical tests are inevitable. For that purpose, we compared a set of quantities computed in two ensembles of lattice configurations:
\begin{enumerate}
\item 
\emph{``Recursion'' lattices} --- independent two-dimensional lattice configurations generated with the probability distribution given by the proposed vacuum wave functional, with $m$ fixed at given $\beta$ to get the correct value of the fundamental string tension $\sigma_\mathrm{F}(\beta)$. The recursion method was described with all details in Ref.\ \cite{Greensite:2007ij}. 
\item	
\emph{Monte Carlo lattices} --- two-dimensional slices of configurations generated by Monte Carlo simulations of the three-dimensional euclidean SU(2) lattice gauge theory with the standard Wilson action; from each configuration, only one (random) slice at fixed euclidean time was taken. 
\end{enumerate}

\subsection{Mass gap}

\begin{figure}
\centerline{\includegraphics[width=0.55\textwidth]{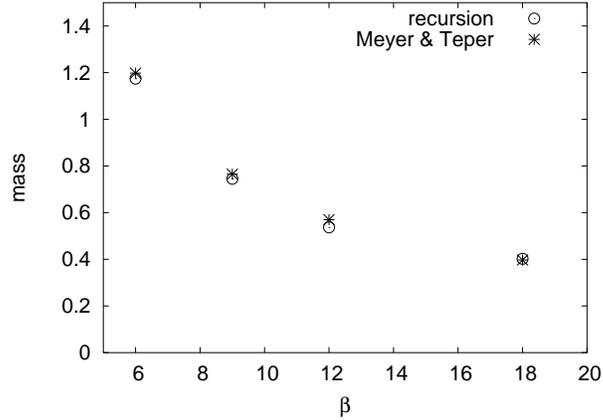}}
\caption{Mass gaps from ensembles of recursion lattices at various lattice couplings, compared to the 0$^{+}$ glueball masses in $D=2+1$ obtained in Ref.\ \cite{Meyer:2003wx} via standard lattice Monte Carlo methods. (From Ref.\ \cite{Greensite:2007ij}.)}\label{fig:mass-gap} 
\end{figure}

	The first such test was performed in Ref.\ \cite{Greensite:2007ij}. We computed the equal-time connected $B^2$--$B^2$ correlator and determined the value of the mass gap from a best fit to its exponential fall-off at large distances. The result for recursion lattices is compared in Fig.\ \ref{fig:mass-gap} to the values of the $0^+$ glueball mass computed in high-statistics simulations of the three-dimensional \YM theory by Meyer and Teper \cite{Meyer:2003wx}. The deviations are at the level of at most 6\%.

\subsection{Coulomb-gauge quantities}

	Another set of quantities of interest are defined in the Coulomb gauge. It was argued by Gribov \cite{Gribov:1977wm} and Zwanziger \cite{Zwanziger:1998ez} that the low-lying spectrum of the \FP operator in Coulomb gauge probes properties of non-abelian gauge fields that are crucial for the confinement mechanism. The ghost propagator in Coulomb gauge and the color-Coulomb potential are directly related to the inverse of the \FP operator, and play a role in various confinement scenarios. In particular, the color-Coulomb potential represents an upper bound on the physical potential between a static quark and antiquark, which means that a confining color-Coulomb potential is a necessary condition to have a confining static quark potential \cite{Zwanziger:2002sh}.
	 
	Our aim \cite{Greensite:2010tm} was to see how well the proposed vacuum wave functional can reproduce the values of Coulomb-gauge observables that can be obtained by standard lattice MC techniques. For that purpose we had to measure them in Coulomb-gauge configurations distributed with the probability following from the square of our temporal-gauge vacuum wave functional. In the operator formalism, the minimal Coulomb gauge is a gauge fixing within the temporal gauge of the remnant local gauge invariance. (See Ref.\ \cite{Greensite:2004ke} for a detailed discussion of this point.) The wave-functional in Coulomb gauge is the restriction of the wave functional in temporal gauge to transverse fields in the fundamental modular region $\Lambda$:
\begin{equation}\label{eq:WFcoul}
\Psi^{\mathrm{Coulomb}}[A_\perp]=\Psi[A_\perp],\qquad A_\perp\in\Lambda.
\end{equation}
	The vacuum expectation value of an operator $Q$ in Coulomb gauge can then be computed from
\begin{equation}
\langle Q\rangle =\langle\Psi_0^{\mathrm{Coulomb}}\vert Q[A_\perp]\vert\Psi_0^{\mathrm{Coulomb}}\rangle
= \langle\Psi_0\vert Q\left[{}^\Omega \! A\right]\vert\Psi_0\rangle,
\label{eq:vev}
\end{equation}
\textit{i.e.}\ we generate configurations following the probability distribution $\Psi_0^2$, transform them to the Coulomb gauge, and evaluate the observable $Q$ in the transformed configuration. ($\Omega$~denotes the gauge transformation that brings the configuration $A$ to the minimal Coulomb gauge.) From the path-integral representation of the vacuum state, we may also go from (\ref{eq:vev}) to
\begin{equation}
\langle Q \rangle = \left\langle Q\left[{}^{\Omega'}\!\! A({\bf x},t=t_0)\right] \right\rangle
\end{equation}
where the right hand side is the expectation value obtained in $D=3$ Euclidean dimensions, and $\Omega'$ is the gauge transformation which takes the gauge field on a $t=t_0$ time-slice into Coulomb gauge.

\begin{figure}
\begin{tabular}{c c}
{\includegraphics[width=0.48\textwidth]{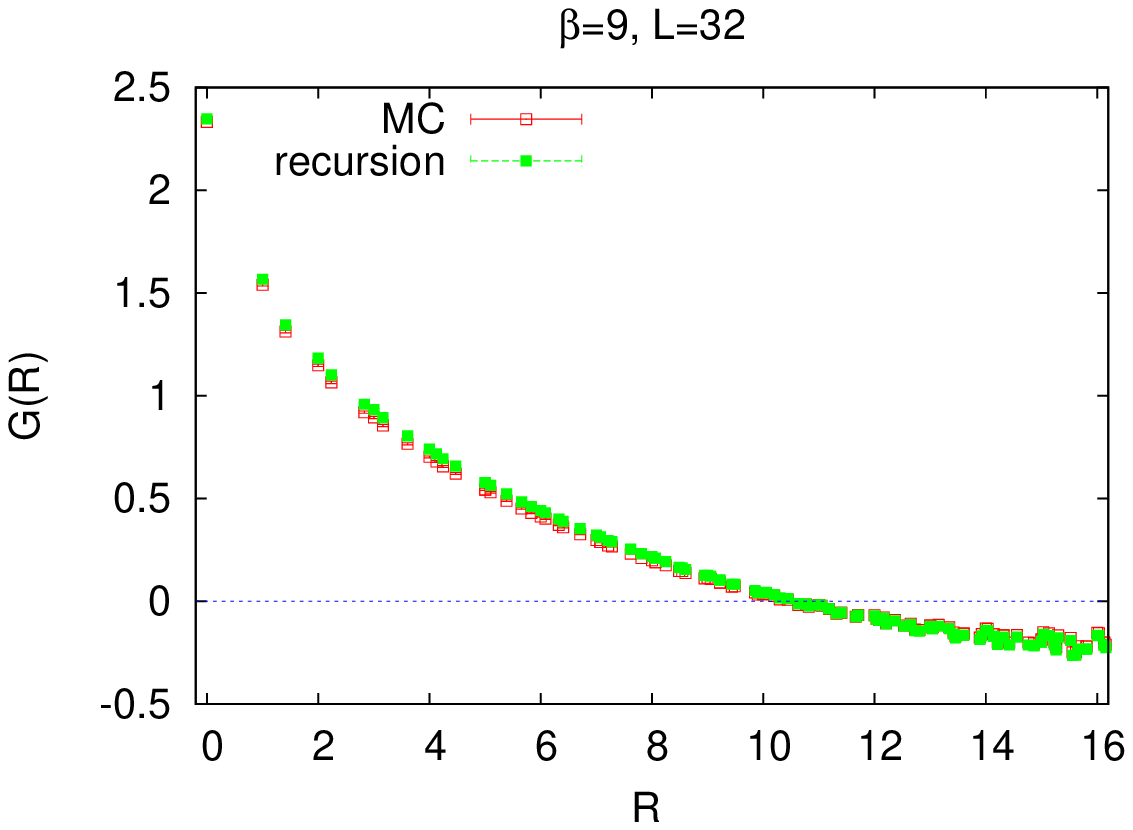}} & {\includegraphics[width=0.48\textwidth]{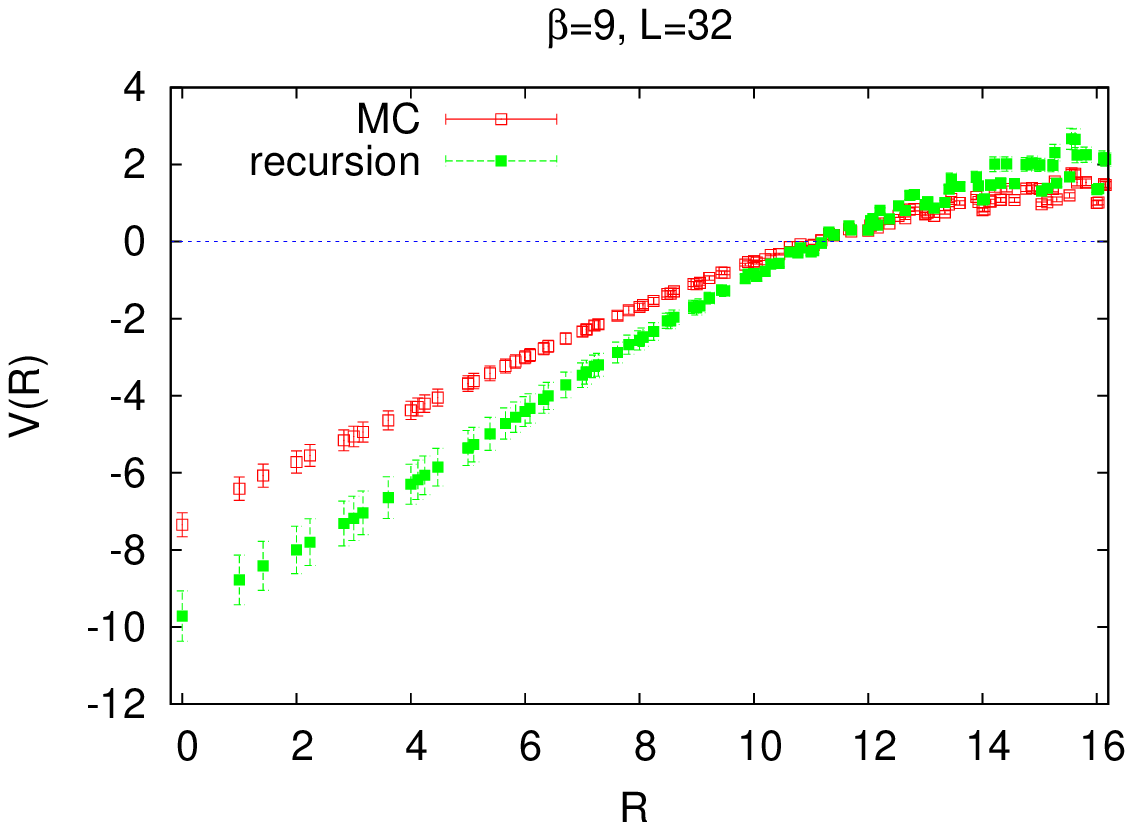}}\\
~~~~~~~~~{\small(a)} & ~~~~~~~~{\small(b)}\\
\end{tabular}
\caption{The Coulomb-gauge ghost propagator (a) and the color-Coulomb potential (b) at $\beta=9$ on $32^2$ lattice.}\label{fig:b9_l32_all}
\end{figure}

	Fig.\ \ref{fig:b9_l32_all} displays results for recursion and Monte Carlo lattices at $\beta=9$ on $32^2$ lattice. The ghost propagator in Coulomb gauge was computed from the inverse of the \FP operator
(in the subspace orthogonal to trivial constant zero modes due to lattice periodicity)
\begin{equation}
G(R)=\left.\left\langle\left({\cal{M}}[A]^{-1}\right)^{aa}_{xy}\right\rangle\right\vert_{\vert x-y\vert=R}
=\left.\left\langle\left(-\frac{1}{\nabla\cdot{\cal{D}}[A]}\right)^{aa}_{xy}\right\rangle\right\vert_{\vert x-y\vert=R},
\label{eq:ghost_prop}
\end{equation}
where
\begin{eqnarray}
\nonumber
  {\cal{M}}^{ab}_{xy}&=& \delta^{ab} \sum_{k=1}^2\left\{ \delta_{xy}  \left[b_k(x)
    + b_k(x-\hat{k})\right] - \delta_{x,y-\hat{k}} b_k(x) - \delta_{y,x-\hat{k}} b_k(y) \right\}\\
    &-& \epsilon^{abc} \sum_{k=1}^3\left\{ \delta_{x,y-\hat{k}} a^c_k(x)
                  - \delta_{y,x-\hat{k}} a^c_k(y)  \right\} 
                  \label{eq:FP}
\end{eqnarray}
with $b_k(x)=\frac{1}{2}\mbox{Tr}[U_k(x)]$ and $a_k^c(x)=\frac{1}{2i}\mbox{Tr}[\sigma^c U_k(x)]$, while the color-Coulomb potential between a static quark and antiquark located at points $x$ and $y$ is proportional to
\begin{equation}
V(R)=\left.-\left\langle\left({\cal{M}}[A]^{-1}(-\nabla^2){\cal{M}}[A]^{-1}\right)^{aa}_{xy}\right\rangle\right\vert_{\vert x-y\vert=R}.
\label{eq:Coulomb_potential}
\end{equation}

\begin{table}[b!]
\centering
\begin{tabular}{p{3cm} p{3cm} p{3cm}}
\hline
\tabentry{range of $\vert V(0)\vert$} & \tabentry{Monte Carlo} & \tabentry{recursion}\\
\hline
\tabentry{$\le$ 10} & \tabentry{867} & \tabentry{786} \\
\tabentry{10 -- 20} &  \tabentry{99} & \tabentry{148} \\
\tabentry{20 -- 100} & \tabentry{32} &  \tabentry{58} \\
\tabentry{100 -- 200} & \tabentry{2} & \tabentry{4} \\
\tabentry{$>$ 200} & \tabentry{0} & \tabentry{2} \\
\hline
\end{tabular}
\caption{Numbers of configurations in different ranges of $\vert V(0)\vert$ at $\beta=9$ on $32^2$ lattice.}\label{tab:mult}
\end{table}

	The agreement of the ghost propagator computed in both sets of lattices is almost perfect [Fig.~\ref{fig:b9_l32_all}(a)], the differences are at the level of the size of symbols. On the other hand, there is a considerable deviation of color-Coulomb potentials computed for the MC ensemble from those for recursion lattices [Fig.\ \ref{fig:b9_l32_all}(b)]. Fortunately, it is not difficult to explain the origin of this deviation. There exist, in both ensembles, ``exceptional'' configurations with a very low (though still positive) lowest nontrivial eigenvalue of the \FP operator. These configurations were extremely difficult to gauge-fix to the Coulomb gauge. If one evaluates the potential in each single configuration, the exceptional ones possess a very high absolute value of the potential at the origin, $\vert V(0)\vert$. One can then classify configurations by their values of $\vert V(0)\vert$, and evaluate average potentials from sets of configurations satisfying a number of cuts: $\{\vert V(0)\vert<\kappa_i, i=1,2,\dots, K\}$. For illustration, Table \ref{tab:mult} lists numbers of configurations in a couple of $\vert V(0)\vert$ bins at $\beta=9$ ($32^2$ lattice). 
	
\begin{figure}
\begin{tabular}{c c}
{\includegraphics[width=0.48\textwidth]{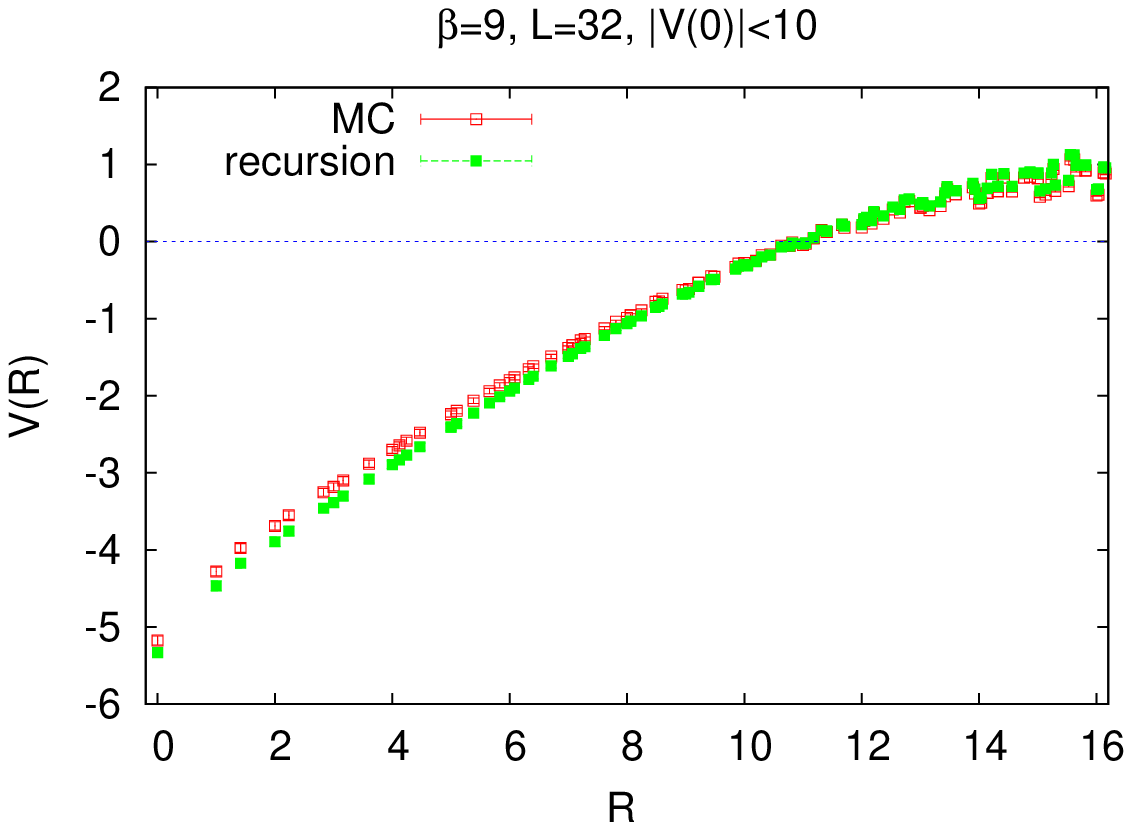}} & {\includegraphics[width=0.48\textwidth]{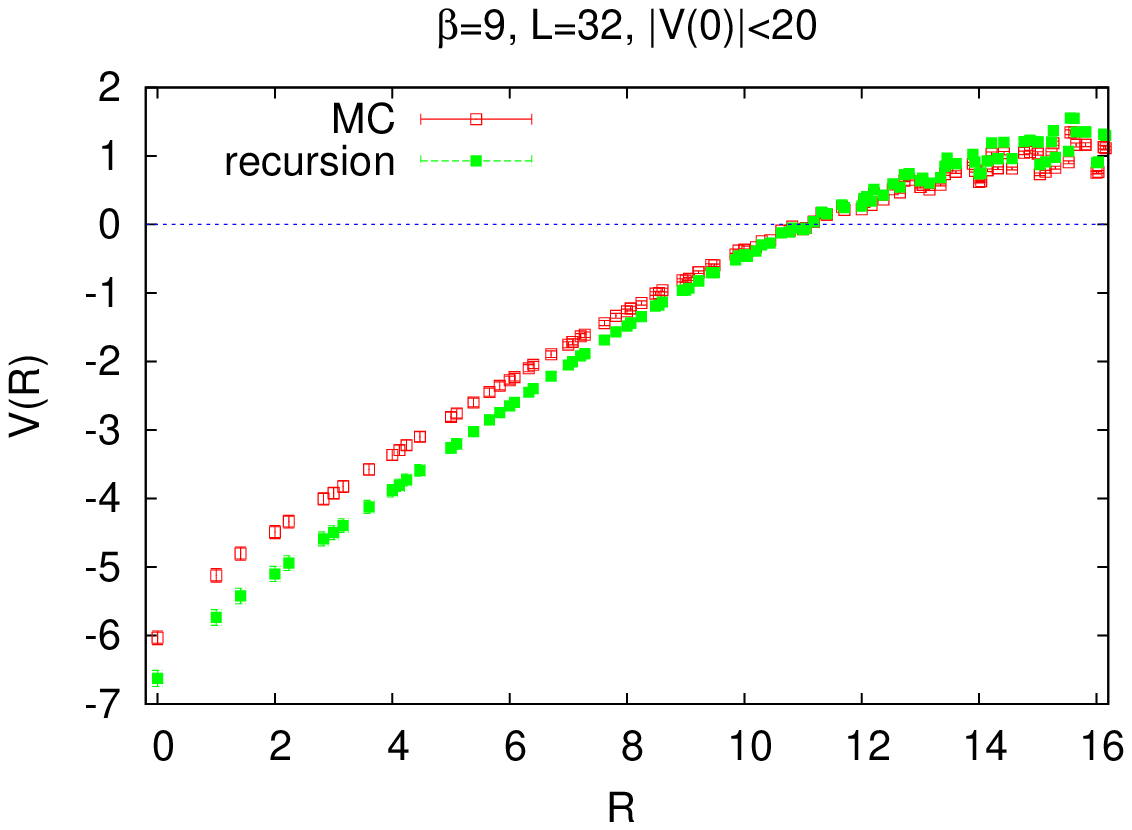}}\\
\end{tabular}
\caption{The color-Coulomb potential for $\kappa=10$ (left) and $\kappa=20$ (right). $\beta=9$, $32^2$ lattice.}\label{fig:b9_l32_cuts}
\end{figure}

\begin{figure}
\begin{tabular}{c c}
{\includegraphics[width=0.48\textwidth]{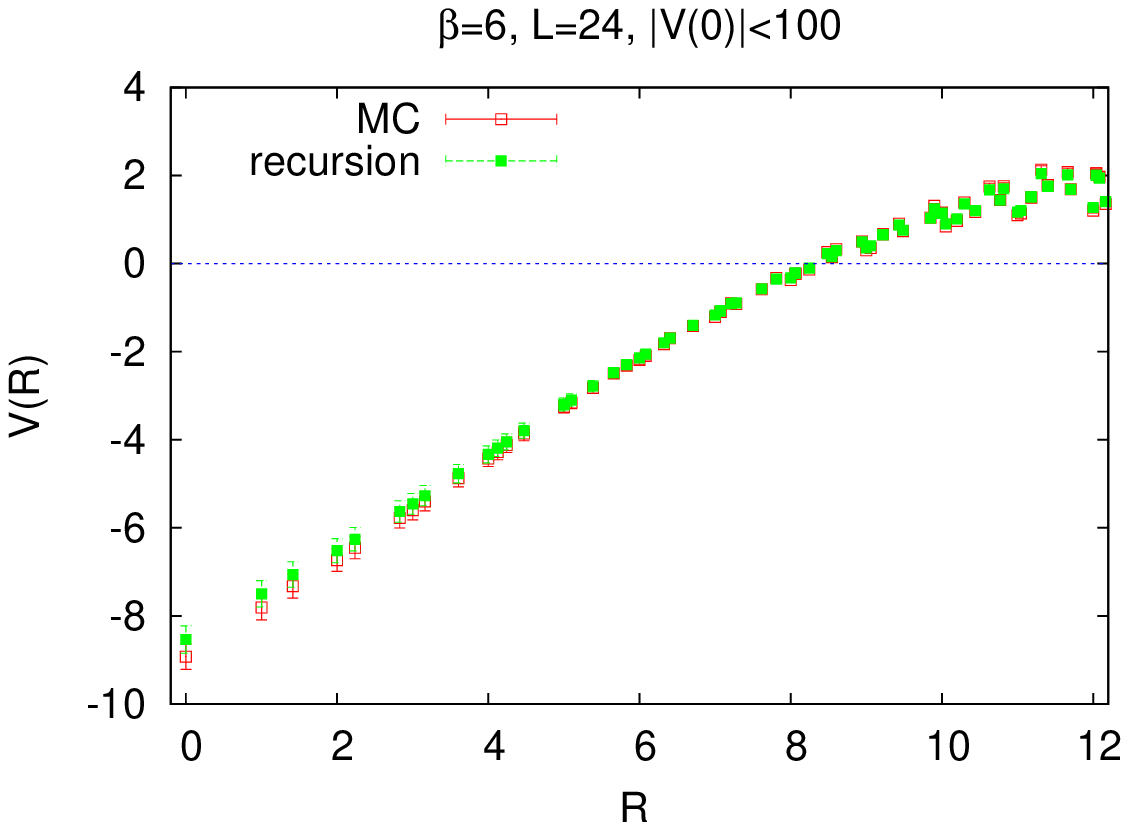}} & {\includegraphics[width=0.48\textwidth]{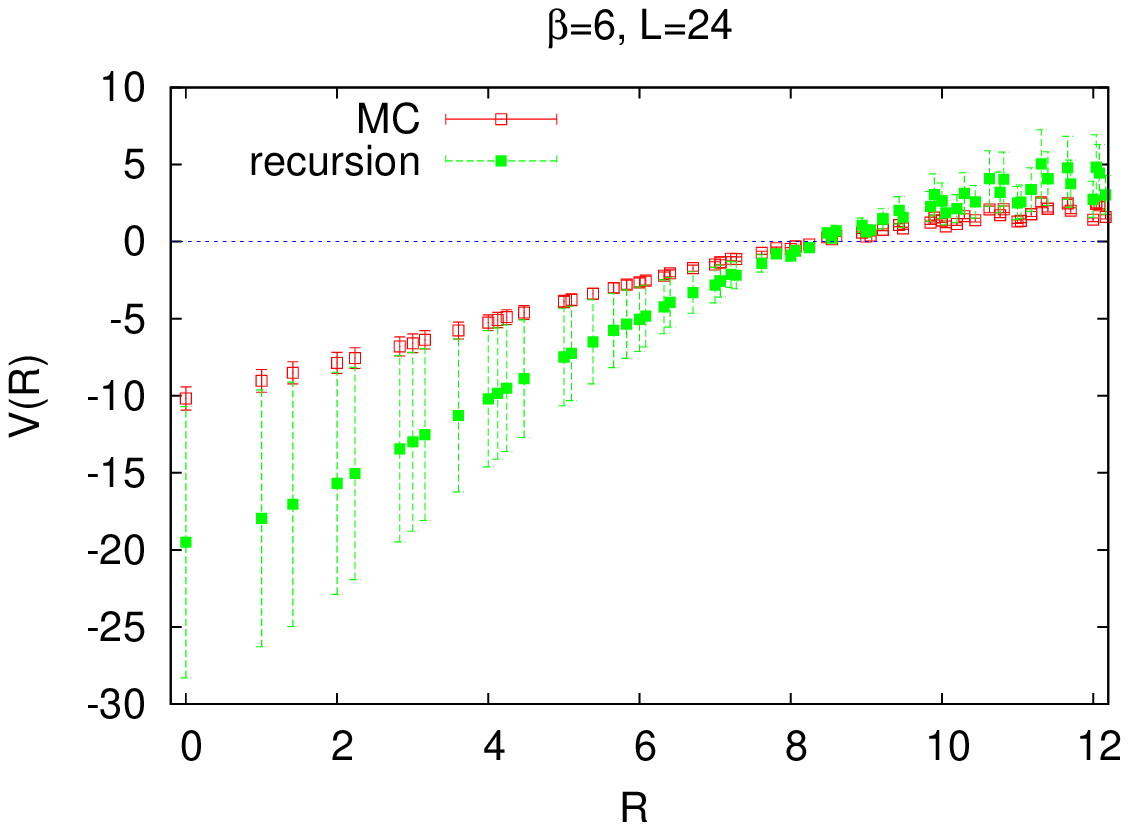}}\\
\end{tabular}
\caption{The color-Coulomb potential for $\kappa=100$ (left) and without cut (right). $\beta=6$, $24^2$ lattice.}\label{fig:b6_l24_cuts}
\end{figure}

	Fig.\ \ref{fig:b9_l32_cuts} shows results for two values of $\kappa$ at $\beta=9$ ($32^2$ lattice). The potentials agree quite well for lower $\kappa$ cut (satisfied by about 80\% lattices), but the differences (and errorbars) grow with increasing $\kappa$. The problem appears even more spectacular at $\beta=6$ ($24^2$ lattice, Fig.\ \ref{fig:b6_l24_cuts}). There is close agreement of color-Coulomb potentials between Monte Carlo and recursion lattices up to the cutoff $\kappa$ as high as 100, but then a single recursion lattice with extremely high value of $\vert V(0)\vert$ completely distorts the picture. Generally, the intervals with high $\vert V(0)\vert$ values are very unequally populated and lead to wild fluctuations of the results, when the cut $\kappa$ is increased. We believe that approximate agreement would be restored with sufficient (but obviously huge) statistics, even though some differences might persist.

\section{Conclusion and outlook}

	The proposed vacuum wave functional for the temporal-gauge SU(2) \YM theory in 2+1 dimensions, Eq.\ (\ref{eq:GO-wf}), seems a fairly good approximation to the true ground state of the theory. 
We have added two new pieces of evidence in its favor:
\begin{enumerate}
\item
The ghost propagator in Coulomb gauge is practically identical in recursion and Monte Carlo ensembles.
\item
With the same statistics of rare exceptional configurations we expect also the color-Coulomb potential from recursion lattices to be close to that determined from Monte Carlo lattices.
\end{enumerate}

	Our further goals are to compare consequences of our proposal to others existing in the literature (see \textit{e.g.\/} \cite{Karabali:1998yq}), to determine the wave functional in numerical simulations for typical field configurations by the method of Ref.\ \cite{Greensite:1988rr}, to improve on the variational estimate of the parameter $m$, and to generalize the proposal to the realistic case of $3+1$ dimensions.

\end{document}